\documentclass[english,aps,superscriptaddress,amsmath,amssymb,longbibliography,onecolumn,floatfix,10pt]{revtex4-1}  
\usepackage{amsmath}
\usepackage{graphicx}  
\usepackage{dcolumn}   
\usepackage{bm}        
\usepackage{amssymb}   
\usepackage[normalem]{ulem}

\usepackage{tikz}
\usetikzlibrary{arrows, arrows.meta}
\usepackage{hyperref}
\hypersetup{
    colorlinks=true,
    linkcolor=blue,
    filecolor=blue,      
    urlcolor=black,
    citecolor=blue,
        }
\usepackage{subfigure}

\begin{document}
\title{Universal Trajectories of Motile Particles  Driven by Chemical Activity}

\author{Chaouqi Misbah}
\email{chaouqi.misbah@univ-grenoble-alpes.fr}
\affiliation{Univ. Grenoble Alpes, CNRS, LIPhy, F-38000 Grenoble, France}
\author{Suhail M. Rizvi}
\affiliation{Univ. Grenoble Alpes, CNRS, LIPhy, F-38000 Grenoble, France}
\affiliation{Current address: Department of Biomedical Engineering, Indian Institute of Technology Hyderabad, Sangareddy, Telangana 502285, India} 
\author{Wei-Fan Hu}
\affiliation{ Department of Mathematics, National Central University, 300 Zhongda Road, Taoyuan 320, Taiwan}
\author{Te-Sheng Lin}
\affiliation{Department of Applied Mathematics, National Chiao Tung University,
1001 Ta Hsueh Road, Hsinchu 300, Taiwan
} 
\author{Salima Rafai}
\affiliation{Univ. Grenoble Alpes, CNRS, LIPhy, F-38000 Grenoble, France}
\author{Alexander Farutin}
\affiliation{Univ. Grenoble Alpes, CNRS, LIPhy, F-38000 Grenoble, France}


\begin{abstract}
 Locomotion is essential  for living cells. It enables bacteria and algae to explore space for food, cancer to spread, and immune system to fight infections. 
 Motile cells display  trajectories of intriguing complexity, from regular (e.g. circular, helical, and so on) to irregular motions (run-tumble), the origin of which has remained elusive for over a century.  This dynamics 
 versatility is conventionally attributed   to the shape asymmetry of the motile entity, to the suspending media, and/or  to  stochastic regulation. We propose here a universal approach highlighting that these movements are generic, occurring  for a large class of cells and artificial microswimmers, without the need of invoking shape asymmetry
nor stochasticity, but are encoded in their inherent nonlinear evolution. 
 We show, in particular, that for a circular and  spherical particle moving in a simple  fluid,  circular, helical and chaotic motions (akin to a persistent random walk) emerge naturally
in different regions of  parameter space. 
This establishes the operating principles for complex trajectories manifestation of motile systems, and offers a new vision  with minimal ingredients. {\color{black} The reduced evolution equations based on symmetries are consistent with those derived   for a model of   an autophoretic particle including diffusion, emission/absorption at the particle surface and hydrodynamics, and provide qualitative and quantitative agreement.}
\end{abstract}
\maketitle

\noindent {\it Introduction.} Cells range in size from a micrometer for a bacterium to a few dozen micrometers for eukaryotic cells. The physics of their movement is thus dominated by the viscosity of water. They have specialized structures, the flagella or cilia, whose movements are maintained by complex biochemical mechanisms. There is also increasing evidence that non flagellated eukaryotic cells that have long been assumed to require a substrate for migration (crawling) can actually swim\cite{barry2010dictyostelium,ONEILL2018,Farutin2019,Theodoly} in a fluid (fluid crawling). One common feature of motile microorganisms is their random-like trajectories sometimes qualified as run-and-tumble \cite{Mori2020,angelani2014,Rupprecht2016,RevModPhys2011,Polin2009} which constitutes a potential optimal way to span the space for survival. 

In the seminal work of Jennings (1901)\cite{Jennings1901}, many microorganisms such as zoospores, flagellate and ciliated Protista are shown to swim in spiral. It is now well established that curved trajectories, such as spiral, circular and helical,  are ubiquitous in nature\cite{Riedel2005,jana2012paramecium,berg1993random,cates2013active,Shenoy2007,Riedel2005,jana2012paramecium} as well as for synthetic non Brownian microswimmers  \cite{kruger2016curling,Loewen2016, suga2018self, narinder2018memory,izri2014self,Hu2019}. The occurrence of such curved trajectories is conventionally retrieved in theoretical models by either invoking stochasticity of the system, chirality of the motile particle, a surrounding non Newtonian fluid or the presence of bounding walls \cite{Shenoy2007,Riedel2005,berg1993random,cates2013active,LaugaBJ,Shenoy2007,Loewen2016,narinder2018memory,izri2014self,Hu2019}. \par

The present goal is to propose a generic theory, based on symmetry arguments, that produce  observed run and tumble trajectories as well as circular, helical and chaotic trajectories. The tour de force here is that these non linear features are obtained for spherical particles moving in an isotropic medium in a deterministic manner. The  symmetry breaking emerges here from the intrinsic non linearities of the problem and thus do not have to be introduced in an {\it ad hoc} manner.

 \noindent {\it Model.} We consider a swimmer powered by a scalar field, say a concentration field $c$ which evolves in space and time.  Marangoni-driven particles \cite{MLB13,morozov2019nonlinear,schmitt2013swimming}, and acto-myosin assisted  cell motility \cite{hawkins2011spontaneous,Voituriez2016,Farutin2019} are two typical examples. In these explicit examples the concentration fields obey advection-diffusion equations, with boundary conditions (such as chemical emission at the particle surface, etc...), which are coupled to hydrodynamics (swimming) or friction (crawling) equations. 

 A common feature of motile systems powered by a chemical field is the occurrence of a spontaneous symmetry-breaking \cite{MLB13,morozov2019nonlinear,schmitt2013swimming,hawkins2011spontaneous,Voituriez2016,Farutin2019} (concentration polarity) leading to autonomous swimming. For the sake of simplicity let us begin with a 2D configuration where the particle has a circular shape with radius unity.

\par
{\color{black}
Before presenting the reduced version of dynamics (in terms of two Fourier modes) based on symmetries,  we would like first to outline how these equations can be  obtained from a full model of an autophoretic particle. 
The model consists \cite{MLB13} of a rigid a particle (taken  to be a   circle with radius $a$), which emits/absorbs a solute that diffuses and is advected by the flow. 
In a reduced form  the model takes the form \cite{MLB13}  

\begin{equation}\label{Eq:advection-diffusion}
\frac{\partial c}{\partial t} + \mathbf{u}\cdot\nabla c = {1\over Pe} \Delta c,\;\;   \Delta \mathbf{u}- {\nabla} p=\mathbf{0},\;\;  \nabla \boldsymbol \cdot \mathbf{u}=0,
\end{equation}
with boundary conditions \begin{equation}\label{BC}
\frac{\partial c}{\partial r}(1,\theta,t) = -{A},\;\;\; \mathbf{u}(1,\theta)  =M\nabla_s c\end{equation}  $\mathbf{u}$  and $p$ is the velocity and pressure  fields, and $\eta$ is the fluid viscosity. 
 P\'{e}clet number is defined by $Pe = |\mathcal{A}\mathcal{M}|a/D^2$, where $D$ is the diffusion constant,    $A = \mathcal{A}/|\mathcal{A}| = \pm 1$ is the dimensionless emission rate ($A>0$: emission, $A<0$: adsorption), and $M=  
 \mathcal{M}/\lvert \mathcal{M} \rvert $ is the dimensionless  particle mobility. Due to the logarithmic divergence of concentration field  in 2D, the size is taken to be finite, so that that  the boundary condition for the concentration field is  $c(R,\theta,t) = 0$ (see discussion in 3D below), 

The velocity field can be expressed as  $\mathbf{u} = (\frac{1}{r}\frac{\partial\psi}{\partial\theta},-\frac{\partial\psi}{\partial r})$ (in polar components), where $\psi$ is the stream function and 
has the following analytical form~\cite{SHHVLT16,Blake1971}
\begin{equation}
\psi(r,\theta,t) = \sum_{\ell=-\infty}^\infty\frac{1-r^2}{2r^{|\ell|}}ikM {c}_\ell (1,t)e^{i\ell\theta},
\label{psi}
\end{equation}
 where we have used the boundary condition $\mathbf{u}(1,\theta)  =M\nabla_s c$ to express the series coefficients in terms of $c$.
Inserting (\ref{psi}) into diffusion equation ( Eq.(\ref{Eq:advection-diffusion})) we obtain a  closed nonlinear equation for $c(\mathbf{r},t)$
\begin{equation}
\label{Eq:advection-diffusion2}
\frac{\partial c(\boldsymbol r,t)}{\partial t  }= \sum _k { c_k(1,t) M e^{ik\theta } \over 2 r^{\vert k\vert +1} }  \left[ (r^2-1)k^2 \frac{\partial c (\boldsymbol r,t) }{\partial r}  +ik(2r^2 +(1-r^2) \vert k\vert ) \frac{\partial c (\boldsymbol r,t)}{\partial \theta} \right]+ {1\over Pe}\Delta c (\boldsymbol r,t) 
\end{equation}

A stationary solution where there is no net flow and zero phoretic velocity  exists at all P\'{e}clet numbers with the solute concentration $c_0(r) = \ln(R/r)$. 
In general $c$ can be written as $c(r, \theta, t)=\sum_{\ell =-\infty}^\infty \hat{c}_\ell (r,t) e^{i\ell \theta}$. 
Neglecting  higher order but linear terms, the following relation for the  Fourier mode $\hat{c}_\ell$ is obtained:
\begin{equation}\label{Eq:LSA}
\frac{\partial {c}_\ell}{\partial t} = -AM \ell^2 \frac{1-r^2}{2r^{|\ell|+2}}{c}_\ell(1, t) + \frac{1}{Pe}\left(\frac{\partial^2}{\partial r^2}+\frac{1}{r}\frac{\partial}{\partial r} - \frac{k^2}{r^2}\right) {c}_\ell (r,t).
=\hat L_\ell(Pe) c_\ell 
\end{equation}
were $\hat L_\ell$ is the linear operator. Looking for solutions in the form $c_\ell=f_\ell (r) e^{i\sigma t}$,  previous calculations\cite{MLB13,Hu2019} showed that $\sigma_\ell \sim (Pe-Pe_{\ell})$, were $ Pe_{\ell}$ is the critical P\'eclet number for which the $\ell$-th harmonic becomes unstable. It has also been shown that the first harmonic ($\ell=1$) becomes first unstable. By increasing $Pe$ the second harmonic becomes also unstable, and so on. 
We assume that $Pe-Pe_{\ell}$ is small close enough, where $P_\ell$ is the critical P\'eclet number for .instability of $\ell$-th harmonic. 
This is just a formal requirement that allows us to keep  dynamics only of these two harmonics. It will be shown that retaining only these two modes is sufficient to exhibit a large panel of behavior going from straight to chaotic motion.

The full equation (\ref{Eq:advection-diffusion2}) can be, by using Fourier decomposition with respect to $\theta$, rewritten as 
\begin{eqnarray}
\label{Eq:advection-diffusion3}
&&\frac{\partial c_\ell ( r,t)}{\partial t  }= \sum _{m\ne \ell}  {m c_m(1,t) M \over 2 r^{\vert m\vert +1} }  \left[ (r^2-1)m \frac{\partial c_{\ell -m}( r,t) }{\partial r}  + (m-\ell)  (2r^2 +(1-r^2) \vert m\vert ) c _{\ell -m} (r,t) \right]+ \hat L_\ell (Pe_{\ell})  c_\ell  ( r,t) \nonumber \\&&+( \hat L_\ell(Pe) -\hat L_\ell (Pe_{\ell})) c_\ell (r,t)  \equiv Q(r,t)+ \hat L_\ell (Pe_{\ell}) c_\ell  ( r,t) \end{eqnarray}

The approach is thus to write 
\begin{equation}
\label{split}
c_l(r,t)=C_l(t)f_{l,0}(r)+\delta c_l(r,t)\,\,\,l\in\{1,2\},
\end{equation}
where $C_l(t)$ is the complex amplitude, $f_{l,0}(r)$ is the proper function such that $\hat L_l(Pe_{\ell,c}) f_{l,0}(r)=0$, and $\delta c_l(r,t)$ is a projection of the function $c_l(r,t)$ on the space of all other proper functions of the operator $\hat L_l(Pe_l)$:
\begin{equation}
\label{deltac}
\delta c_l(r,t)=\sum\limits_{k>0}C_{l,k}(t)f_{l,k}(r),\,\,\,\hat L_l(Pe_l)f_{l,k}(r)=\lambda_{l,k}f_{l,k}(r).
\end{equation}
Here the functions $f_{l,k}(r)$ are the proper functions of the operator $\hat L_l(Pe_{\ell})$ corresponding to eigenvalues $\lambda_{l,k}$.
Equation(\ref{Eq:advection-diffusion3}) becomes then
\begin{equation}
\label{splitequation}
\partial_t C_\ell f_{\ell,0}(r)=\hat L_l(Pe_{\ell}) \delta c_l(r,t)+\tilde Q(r,t)-\partial_t \delta c_\ell(r,t).
\end{equation}
where $\tilde Q$ is obtained from $Q$ in which we substitute 
$c_\ell(r,t)$ by $C_\ell(t)f_{l,0}(r)+\delta c_\ell(r,t)$

The goal now is to obtain  closed equations for $C_1(t)$ and $C_2(t)$. Close enough to bifurcation these amplitude are small,     $C_1=O(\epsilon)$ and $C_2=O(\epsilon)$, where $\epsilon$, measures distance from criticality, $|Pe-Pe_{\ell}|=O(\epsilon)$. Since the growth rate is small, $\sigma_\ell\sim  |Pe-Pe_{\ell}|=O(\epsilon)$, we also have  
 $\partial_t C_\ell =O(\epsilon^2)$ (usual Landau critical slowing down). Higher order harmonics  amplitudes are smaller; for $l>2$, we have $c_{l}(r,t)=O(\epsilon^{\lceil l/2\rceil})$ and  $\partial_t c_{l}(r,t)=O(\epsilon^{\lceil l/2\rceil+1})$. This implies 
  $\delta c_\ell(r,t)=O(\epsilon^2)$ and $\partial_t\delta c_\ell(r,t)=O(\epsilon^3)$.   $\delta c_\ell(r,t)$  induces cubic terms, meaning that it does not enter the evolution equations of $C_\ell$ to order $\epsilon^2$. The evolution equations can be readily obtained tanks to the Fredholm alternative theorem. For that we need to determine  the kernel of the adjoint operator  $\hat L^+_l(Pe_{\ell})$ : \begin{equation}
\hat L_l^+(Pe_l) g_{\ell ,0}(r)=0,
\end{equation}
The adjoint operator is defined with respect to the inner product
\begin{equation}
\label{inner}
\langle f,g\rangle=\int\limits_1^R f(r)g(r)^* rdr,
\end{equation}
which is chosen to maintain the self-adjoint property of the diffusion operators $\hat D_l$ subject to the boundary conditions of the functions $c_l(r)$.  
By construction  $\delta c_\ell$ satisfies $<g_{\ell,0},  \delta c_\ell>=0$.

By projecting Eq. (\ref{splitequation}) on $g_{\ell,0}$ function we obtain 
the desired equation for $C_\ell$, which formally reads
\begin{equation}\label{C1C2}
\partial_t C_\ell={\langle \tilde Q(r,t),g_{\ell,0} \rangle \over \langle f_{\ell,0} ,g_{\ell,0} \rangle } 
\end{equation}
Using $c_l(r,t)=C_l(t)f_{l,0}(r)$ and collecting in $\tilde Q$  terms involving $C_1(t)$ and $C_2(t)$ (higher harmonics are supposed to be stable)
we straightforwardly obtain the form of the evolution equations for $C_1(t)$ and $C_2(t)$ (which factor out of scalar products) to quadratic order, where coefficients are scalar products involving $f_{\ell,0}(r)$  and $g_{\ell,0}(r)$. These functions can even be determined and the scalar products can be evaluated analytically (see \cite{FarutinC1C2}; since our goal here is to prove the form of the equations, the values of coefficients are unimportant for our purposes). 
 \begin{eqnarray} && \dot C_1 =\sigma_1 C_1 +\beta _1 C_1^* C_2 \nonumber \\ 
&&  \dot C_2  =\sigma_2 C_2 -\beta _2 C_1^2 
\label{C12p}
 \end{eqnarray} 
 It turns out  out that the cubic terms are necessary for the nonlinear saturation, this is why $\delta c_\ell(r,t)$ must be taken into account, and we obtain \cite{FarutinC1C2} 
 \begin{eqnarray} && \dot C_1 =\sigma_1 C_1 +\beta _1 C_1^* C_2  -  \gamma_1 | C_2|  ^2 C_1 \label{C12p1}\nonumber \\
&&  \dot C_2  =\sigma_2 C_2 -\beta _2 C_1^2  -\gamma_2  | C_1|  ^2 C_2  -  \xi_2 | C_2|  ^2 C_2
\label{C12pp}
 \end{eqnarray} 
 The various coefficients obtained for the phoretic model  are given in \cite{FarutinC1C2}. 
 
}

{\it Nonlinear evolution equations from symmetries.}
{\color{black} The forms of the above set of equations  (\ref{C12pp}) are in fact quite general and do not depend on the explicit model (only the values of the coefficients depend on the model).} 

 The form of nonlinear evolution equations  
must comply with space symmetry. Indeed, due to particle and medium  isotropy, a displacement along the bead periphery (rotation by a certain angle) by a constant amount $\theta_0$  should leave the evolution equations invariant.   The first Fourier mode reads as ${c}_1\sim C_1(t)e^{i\theta }+c.c.$. If  one changes $\theta$ by a constant $\theta_0$ we have   \begin{equation} \theta\rightarrow \theta+\theta_0 \implies C_1\rightarrow C_1e^{i\theta_0}
 \end{equation} 
 Rotation by  constant angle  is equivalent to a phase shift of $c_1$.
  
  More generally, since the evolution equations contains other harmonics, the evolution equations for $C_m(t)$ must be invariant under the transformation 
 
 \begin{equation}  C_m \rightarrow C_m e ^{i m \theta_0}
 \label{inv}
 \end{equation} 
 It is easy to check that the above se of equations (\ref{C12pp})  is invariant under transformation (\ref{inv}). 
 Note that terms of the form $C_1^2$, $C_2^2$, for example, are not eligible, since they do not comply with the symmetry constraints. 
The set (\ref{C12pp}) is thus general and should be expected for any model driven by chemical activity (phoretic models, especially the one dealt with here, and motility driven by acto-myosin are two typical examples). It is also essential to note that the symmetry arguments do not depend on whether the particle is a swimmer, or crawler. {\color{black} For acto-myosin systems the appropriate dimensionless number is a dimensionless myosin contractility $\chi '$ (proportional to myosin contractility divided by viscosity and myosin diffusion) \cite{Farutin2019}, instead of $Pe$. In that case motility takes place for $\chi '$ of order few unities, in consistent with experiments \cite{Farutin2019}.}

 {\color{black} Due to the generic character  of the equations for $C_1$ and $C_2$  we will below look at the model as general, and potentially applicable to various swimmers powered by a concentration field. Differences between systems will only show up in the values of the coefficients. In what follows we will exploit this generality without specific values of coefficients. Our goal is to establish what are the main features exhibited by Eqs. (\ref{C12pp}) by  exploring different values of the coefficients). In Ref.\cite{FarutinC1C2} we can find the expressions of the coefficients for the above phoretic model, and where we show that the reduced model in terms of $C_1$ and $C_2$ captures both qualitatively and quantitatively the results of the full model.} 
 

{\color{black} Above we assumed that $C_2=O(\epsilon )$ for a formal expansion in terms of $\epsilon$. However, the set of equations (\ref{C12pp}) remain valid even if $C_2$ is of smaller order provided we decide to keep both $C_1$ and $C_2$ in the expansion without imposing the $\epsilon$ scaling. For example if $C_2$ is a sufficiently stable mode, its amplitude will be small. This happens if $\sigma_2$ is sufficiently negative, meaning that $C_2$ decays sufficiently fast} to its  
 its steady state value. Using thus the adiabatic elimination of $C_2$, (i.e. $\dot C_2\simeq 0$) one obtains
 $ C_2 =\beta _2 C_1^2/\sigma_2$, and plugging it into 
 (\ref{C12pp}) one obtains to leading order (cubic terms containg $C_2$ are of higher order)
  \begin{equation}  \dot C_1  =\sigma_1C_1 - \alpha_2 | C_1|  ^2 C_1
  \label{C10}
 \end{equation} 
with $\alpha_2=-\beta_2\beta_1/\sigma_2>0$ (supercritical bifurcation; subcritical bifurcations are not considered here). The nonlinear term is stabilizing and leads to saturation of the linear growth. Since $\sigma_2<0$, $\beta_1$ and $\beta_2$ must have the same sign (taken arbitrarily to be positive here; their sign is arbitrary, see below).

Since close to instability $\sigma _1\sim (Pe-Pe_1)$, Eq. (\ref{C10}) has a steady state solution $C_1^{(0)}\sim (Pe-Pe_1)^{1/2}$, and the swimming speed (which is a linear function of $C_1$; see below) behaves in the same way with $Pe$:
  \begin{equation}  v\equiv v^0\sim (Pe-Pe_1)^{1/2}  \label{C2p}
 \end{equation}   The swimmer trajectory is straight. This is a signature of a supercritical bifurcation from non motile to motile state.  This is in qualitative agreement with the numerical finding \cite{MLB13,Hu2019}. 

{\color{black} When $\sigma_2$ becomes sufficiently small (of order $\epsilon$), the second harmonic amplitude $C_2$ becomes comparable to $C_1$, and dynamics is described by the full set (\ref{C12pp})}. 
 Note that  a change  $C_2 \rightarrow -C_2$ (corresponding to a phase shift of $C_2$ by $\pi$)   leads to a simultaneous change of sign of $\beta_1$ and $\beta_2$, this is why  their sign is unimportant. The signs of cubic terms are dictated by the fact that nonlinear terms should saturate the linear growth of instability. That is $\gamma_1,\gamma_2, \xi_2>0$. 
 It is always possible to set, for example $\xi_2$,  coefficient to unity upon an appropriate rescaling. $\sigma_1$ and $\sigma_2$ change sign at two different critical values of $Pe$, denoted as $Pe_1$ and $Pe_2$. 
  
{\it Results.} Setting $C_1 =\rho_1(t) e^{i\phi_1(t)}$ and $C_2 =\rho_2(t) e^{i\phi_2(t)}$ into (\ref{C12pp}) one  obtains
 \begin{subequations} \begin{eqnarray} && \dot \rho_1 =\sigma_1 \rho_1 +\beta _1 \rho_1 \rho_2\cos(\Psi )-  \gamma_1  \rho_2  ^2 \rho_1 \label{eqa}\\
&&  \dot \rho_2  =\sigma_2 \rho_2 -\beta _2 \rho_1^2 \cos(\Psi )-\gamma_2  \rho_1  ^2 \rho_2  -   \rho _2^3 \label{eqb}\\
\rho_2&& \dot\Psi= (\beta_2\rho_1 ^2 - 2 \beta_1 \rho_2 ^2) \sin (\Psi ) , \;\; \; \Psi\equiv \phi_2 - 2\phi_1
 \label{eqc}
  \end{eqnarray}
  \end{subequations} 
The phases $\phi_1$ and $\phi_2$ are determined in terms of $\Psi$, $\rho_1$ and $\rho_2$.  For example, $\phi_1$ obeys 

\begin{equation}
\dot \phi_1 = \beta_1 \rho_2 \sin (\Psi).
\label{pi1}
 \end{equation} That only a single phase  ($\Psi$) matters 
 is a result of rotational invariance.  The phase $\Psi$ plays an important role in the occurrence of curved trajectory, and especially the circular one, which can be handled fully analytically. 
Since $c(s,t)=c_0+\rho_1 e ^{is+i\phi_1 }+\rho_2 e ^{2is+i\phi_2 }+c.c.$ (in what follows $c_0$ will be omitted) 
we can write
 \begin{eqnarray}
 c(s,t)&=&2\rho_1\cos(s+\phi_1) + 2\rho_2\cos(2st +2\phi_1)\cos(\Psi) \nonumber \\
&& -2\rho_2 \sin(\Psi )\sin(2s +2\phi_1) \label{cdrift} \end{eqnarray} 
It is clearly seen that as soon as $\Psi\ne 0, \pi$ the concentration field loses its axial symmetry (generated by second harmonic), which results into a curved trajectory, as shown below. The existence of a non trivial fixed point for $\Psi$ (Eq. (\ref{eqc})) requires
\begin{equation}
\beta_2\rho_1 ^2 - 2 \beta_1 \rho_2 ^2=0
\label{cond}
\end{equation}
 This means that $\beta_1$ and $\beta_2$ must have the same sign (as already discussed before). Setting $\dot\rho_1=\dot \rho_2=0$
in (\ref{eqa})-(\ref{eqb}) determines $\rho_1$ and $\rho_2$  as a function of $\cos(\Psi )$, and using  (\ref{eqc}) leads to  a closed equation for $\Psi$. {The explicit condition for fixed point of $\Psi$ in Eq.(\ref{cond}), which relates  the coefficients $\beta_i$, $\gamma_i$... entering the model, is given in \cite{SI}}. 
 A non trivial fixed point of (\ref{eqc}), yields from  Eq.(\ref{pi1})  $\dot \phi_1= \beta_1 \rho_2^0 \sin (\Psi^0)$  ($'0'$ refers to the fixed point solution). This entails that \begin{equation} 
 \phi_1=\beta_1 \rho_2^0 \sin (\Psi^0) t \equiv v_d t
 \label{vd}
\end{equation} 
 From  Eq. (\ref{cdrift}) we see that the concentration field drifts in time sideways along the bead surface with velocity $v_d$. The bead velocity ${\mathbf v}(t)$ is related to $c(s,t)$ and possibly its gradients. For example, for phoretic particles\cite{morozov2019nonlinear}  $\mathbf{v}(t) \sim  \int_0^{2\pi} \nabla_s c(s,t) ds \sim  [-Re(C_1), Im (C_1)]$. This expression holds  whenever $\mathbf v$ is a linear function of $c$ and its gradients (see another example \cite{Farutin2019}).  The bead Cartesian coordinates are then given by 
 \begin{equation}
 x\sim  {\rho_1^0\over v_d} \sin (v_dt) , \;\; y\sim  - {\rho_1^0\over v_d} \cos (v_dt)
 \label{xpos}
 \end{equation} 
  This is the equation of a circle
   with radius 
 $ R_{circle} \sim   {\rho_1^0/v_d} $. 
 
\begin{figure}
\centering
\includegraphics[width=0.8\textwidth]{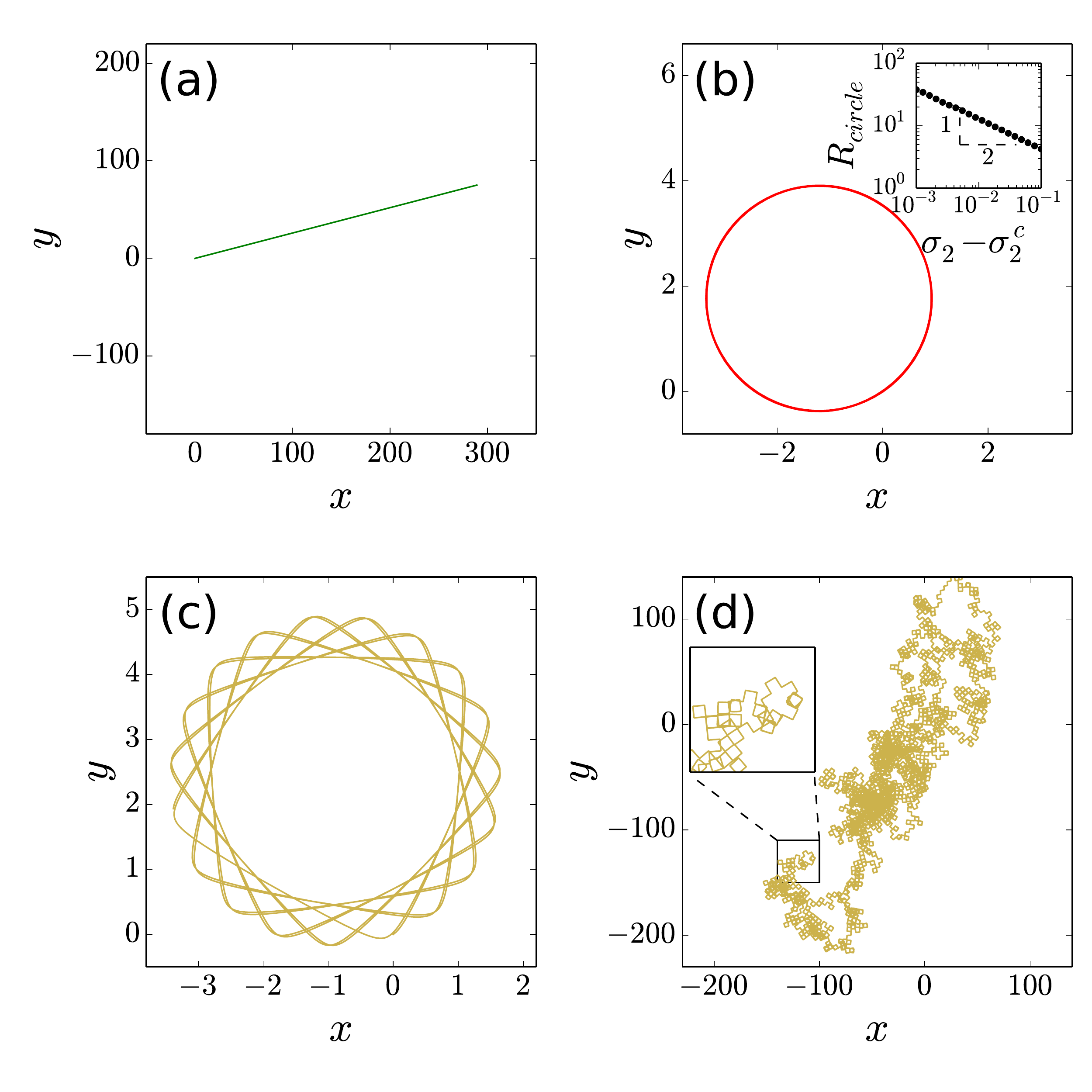}
\caption{\footnotesize Swimming patterns as $\sigma_2$ increases (growth rate of second harmonic). (a) Straight  ($\sigma_2=-1$), (b) circular ($\sigma_2=0.1$; inset shows scaling of circle radius close to criticality confirming prediction (Eq. (\ref{rcircle})) , (c) precession ($\sigma_2=0.4$), and (d) chaotic trajectories ($\sigma_2=1.2$).
\label{swimpattern}}
\end{figure}

One can express  from Eq.  (\ref{eqa})-(\ref{eqb}) $\rho_1^0$ and $\rho_2^0$ as a function of $\cos(\Psi )$, and plugging this into Eq.(\ref{eqc}) we obtain   $\dot \Psi = G[\cos(\Psi )] \sin (\Psi)$, where the function $G$ is listed in \cite{SI}. 
In the vicinity of the emergence of the circular trajectory $G$ is small, 
and we obtain to leading order (see \cite{SI})
\begin{equation} 
 \dot \Psi= {(\sigma_2-\sigma_2^c)\over b} \Psi - \xi \Psi ^3
\end{equation}  
with $b$, $\xi$ positive quantities,   functions of the parameters entering (\ref{eqa})-(\ref{eqc}) (see \cite{SI}). Setting, for example, all coefficients to unity, except $\sigma_2$, serving as a single control parameter, we find $\sigma_{2}^c\simeq -0.09$ (and $b\simeq 0.38$, $\xi\simeq 7.7$), meaning that at criticality the second harmonic is almost neutral (small growth rate).
The bifurcation to circular trajectory is of supercritical nature, and the phase $\Psi^0$ behaves close to the critical point as $(\sigma_2-\sigma{_2^c})^{1/2}$. According to (\ref{vd}) and (\ref{xpos}),   the radius of the circle diverges at the critical point of circular trajectory
\begin{equation}
R_{circle} \sim {1\over  (\sigma_2-\sigma_2^c)^{1/2}}
\label{rcircle} 
\end{equation}
Numerical solution of  (\ref{eqa})-(\ref{eqc}) confirms this prediction (Fig. \ref{swimpattern}). This scaling is observed for  light-sensitive colloidal swimmers  \cite{Narinder2018}, and is expected to be generic.
Numerical solution of   (\ref{eqa})-(\ref{eqc}) reveals a complex dynamics ranging from straight trajectories to chaotic ones. 
To illustrate this,  we have set all parameters to unity, and varied $\sigma_2$ from negative to positive (by keeping  $\sigma_2$ small enough  for higher order harmonics to  play a minor role). 

\begin{figure}
\centering
\includegraphics[width = \columnwidth]{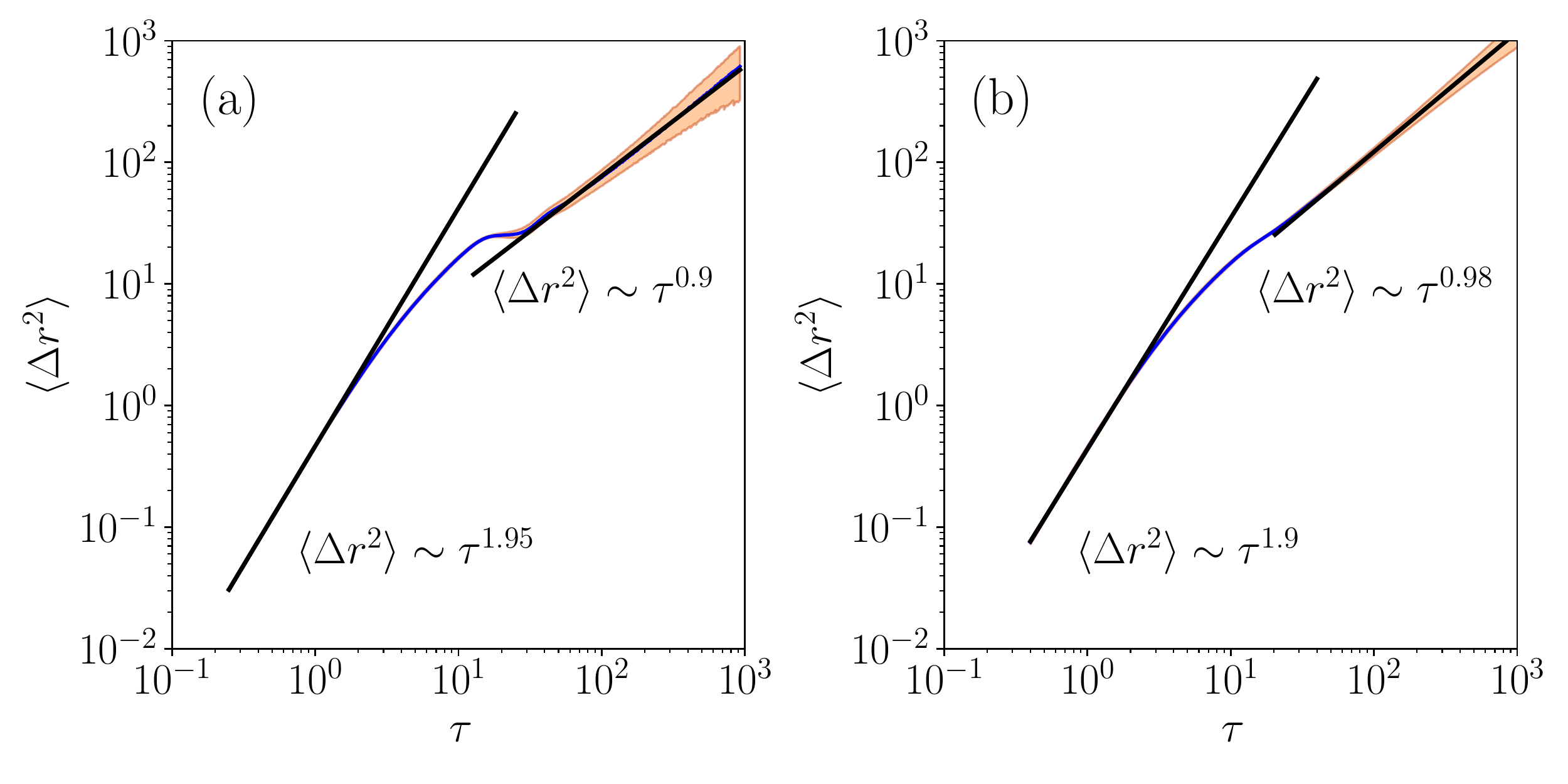}
\caption{\footnotesize Mean square displacement for (a) $\sigma_2=1.0$ and (b) $\sigma_2=1.2$. The shaded region corresponds to the standard deviation in MSD.
\label{chaotic_MSD}}
\end{figure}

Figure \ref{swimpattern} shows a typical swimming pattern, going from straight, circular, precession, to an apparently erratic motion. This last motion (chaos) bears strong resemblance with the run-tumble dynamics, having  a persistent random walk feature. 
We measure the mean square displacement 
\[
\mbox{MSD}(\tau) = \langle \Delta x^2(\tau) \rangle = \langle \|\mathbf{x}(t+\tau)-\mathbf{x}(t)\|^2 \rangle, 
\]
where $\mathbf{x}(t)$ is the location of the particle at time $t$ and $\langle\cdot\rangle$ denotes the average along the entire trajectory. Figure \ref{chaotic_MSD} reports the MSD. At short time we have a ballistic motion, whereas for longer times, a de-correlation process due to chaotic turns of velocity direction leads to a MSD proportional to  ${\tau}^{\kappa}$, where $\kappa$ depends on model parameters, yielding both diffusive and sub-diffusive regimes (Fig. \ref{chaotic_MSD} ). Actulally, it is not obvious that a chaotic motion is equivalent (at long time) to normal diffusion. There are several chaotic maps yielding anomalous diffusion \cite{Geisel}.

{\it Extension to 3D.} As in 2D the 3D model  relies only on two harmonics of the concentration field, $c_i$ and  $c_{ij}$, where $c_i$ is a 3D vector and $c_{ij}$ is a 3D symmetric traceless tensor.
The particle is taken as a unit sphere with a concentration 
\begin{equation}
\label{concentration}
c(\boldsymbol{r})=c_ir_i+c_{ij}r_ir_j.
\end{equation}
where $r_i$ is the ith component of position vector on the particle surface. We propose the following system  (see Supplemental Materials):
\begin{subequations}
\begin{align}
\label{dynamicsa}
	\dot c_i&=\sigma_1 c_i+\alpha_1 c_j^2c_i+\beta_1(c_k^2c_{ij}c_j-c_jc_kc_{jk}c_i)\\
\label{dynamicsb}
	\dot c_{ij}&=\sigma_2 c_{ij}+\beta_2(c_ic_j-\delta_{ij}c_k^2/3).
\end{align}
\end{subequations}
All terms written above are consistent with symmetry (three rotations). 
System (\ref{dynamicsa},b) 
leaves the evolution of the norm of $c_i$ independent of $c_{ij}$ 
\begin{equation}
\label{norm}
	c_i\dot c_i=\sigma_1 c_i^2+\alpha_1 (c_i^2)^2.
\end{equation}
This choice is not necessary, but allows for a complete analytical handling (results are unaffected by this choice (see below).
This is the classical form of a pitchfork bifurcation. We assume  $\alpha_1<0$ (supercritical bifurcation).
With this choice, we obtain that for $\sigma_1<0$ the stable solution is $c_i^2=0$, which corresponds to a non-motile case.
For $\sigma_1>0$, the stable solution is 
\begin{equation}
\label{c10}
	c_i^2=-\sigma_2/\alpha_1,
\end{equation}
which corresponds to a motile solution (recall that the swimming speed $v_i$ is proportional to $c_i$).
Since the norm dynamics of $c_i$ is decoupled, we assume below that the norm of $c_i$ has already reached its stationary value defined by Eq. (\ref{c10}) (in 2D we have actually shown that for circular motion amplitudes $\rho_1$ and $\rho_2$ are constants). As we have seen in 2D a circular trajectory leads to a fixed concentration spot moving along the particle periphery. Instead of using the dynamics of phases ($\phi_1$ and $\phi_2$) as in 2D, we find it more convenient in 3D to follow another approach. The idea is to find if there is a co-rotating frame in which the concentration spot would be steady. Rotation of a spot along the sphere requires some symmetry-breaking. For the straight motion (say along $x$) the spot possesses axial symmetry around that axis. A first obvious breaking of this symmetry leaves  
 a single mirror of symmetry   containing $x$-axis; we take it to be   $x-y$ plane. We will see that the concentration spot will spontaneously move along the equator, and the particle will follow a circular path. The next broken  symmetry is the $x-y$ mirror, which will make the spot to move along a closed trajectory, distinct from  equator, and the particle follows a helical path. It is convenient to solve our system (\ref{dynamicsa})-(\ref{dynamicsb}) in the co-moving frame with angular velocity ${\boldsymbol \omega}$ (to be determined) of concentration spot. The left hand side of (\ref{dynamicsa})-(\ref{dynamicsb}) become (see \cite{SI}) $\dot c_i+ \varepsilon_{ijk} \omega_j c_k$ ($\varepsilon$ is Levi-Civita symbol) and $\dot c_{ij}+ \varepsilon_{ikl} \omega_k c_{kj}+\varepsilon_{jkl} \omega_k c_{ki}$. Then setting 
 $\omega _ i = \beta_1 \varepsilon_{ijk} c_j c_{kl} c_l$ cancels $\beta_1$ term in  (\ref{dynamicsa}), and we are then left with equation of $c_{ij}$ only. 
 
\begin{figure}
\centering
\includegraphics[width = 0.7\columnwidth]{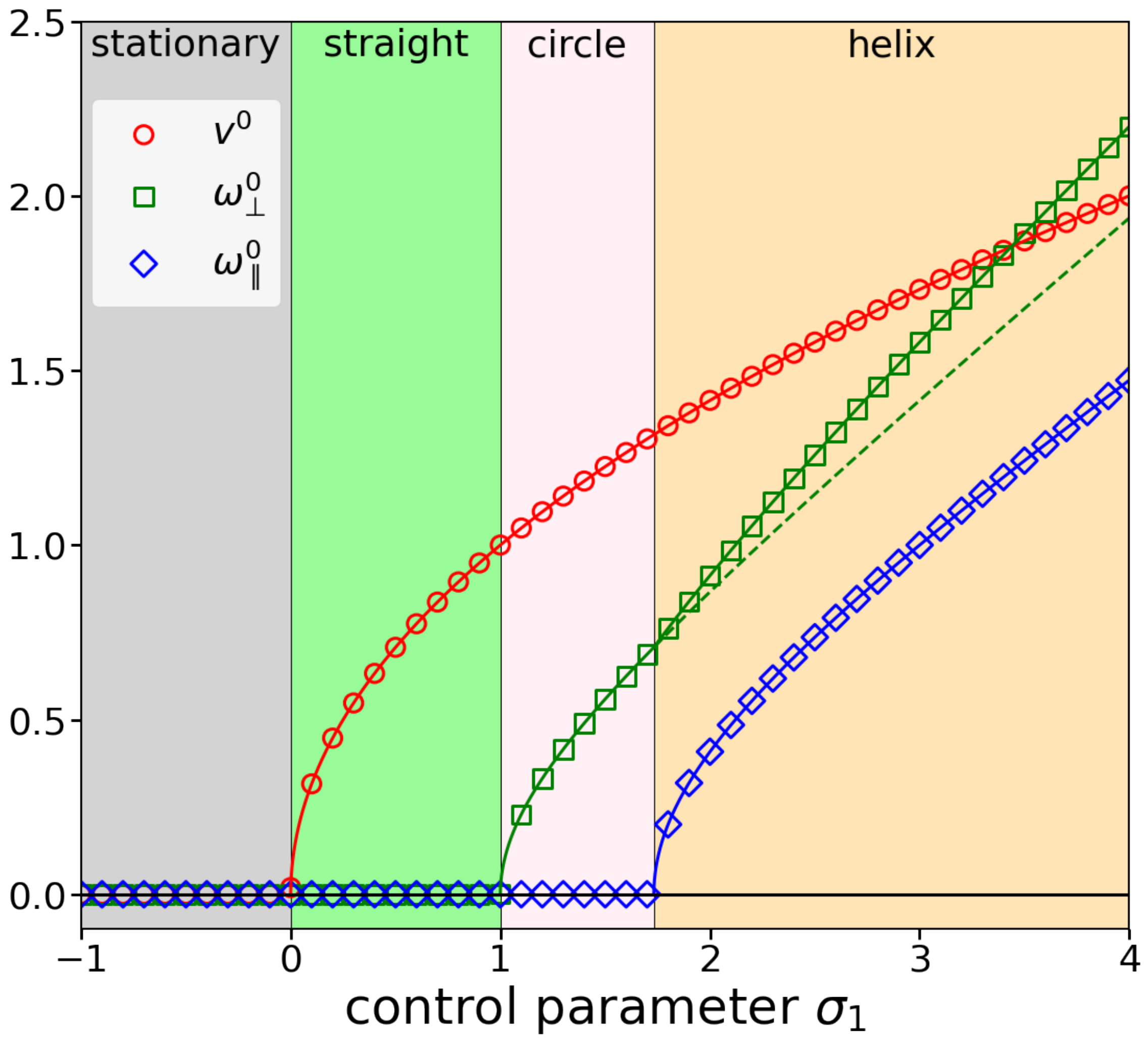}
\includegraphics[width = 0.7\columnwidth]{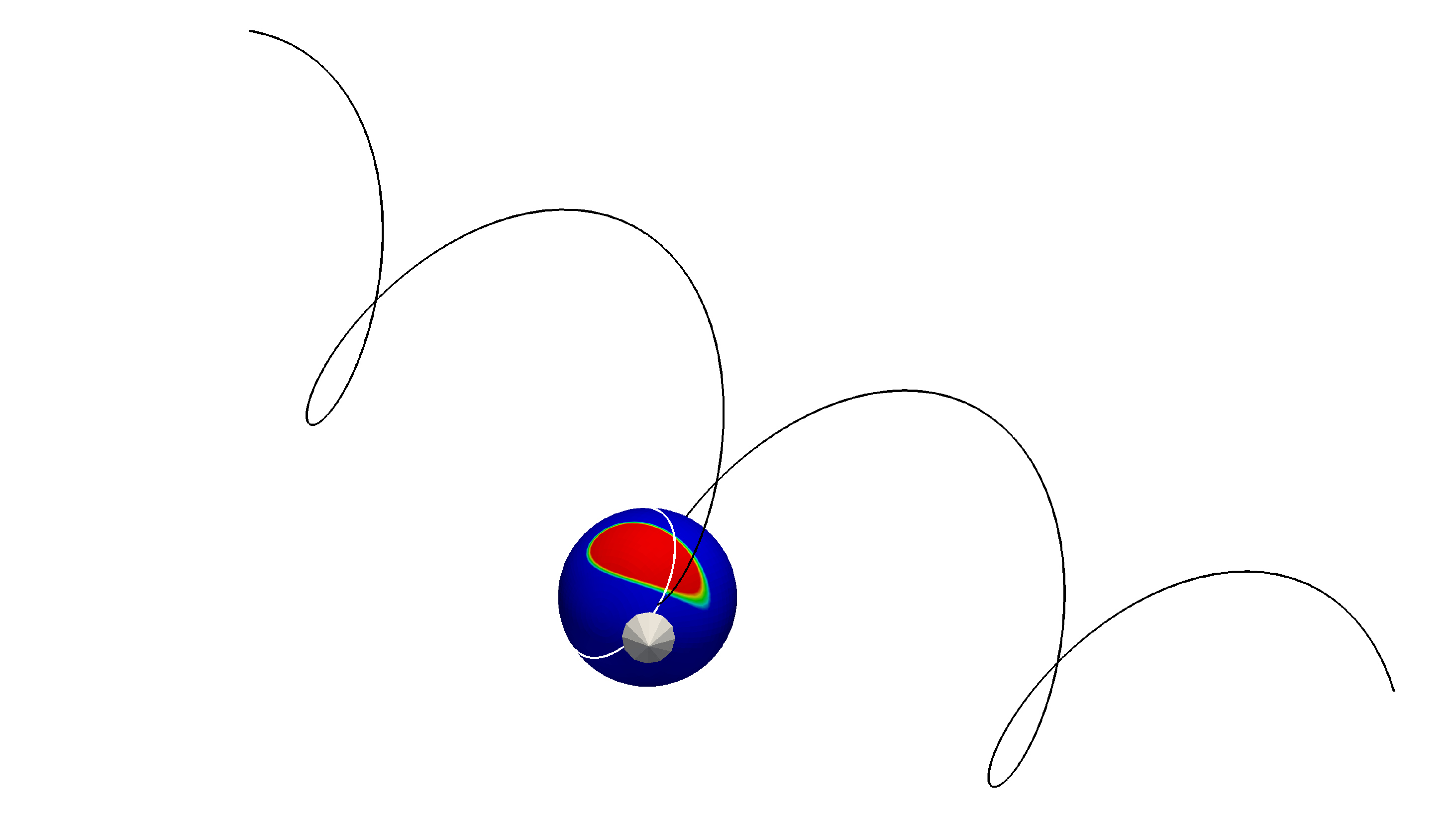}
\caption{Top: Bead velocity showing series of bifurcations from straight, circular to helical motion.The components of angular velocity along the velocity and orthogonal to it are shown as $\omega_\parallel^0$ and $\omega_\bot^0$, respectively. Solid lines refer to analytical solution and symbols to numerical ones. Bottom: Bead trajectory showing helical path and a concentration spot following a circle (white)  outside the equator.
\label{scheme3D}}
\end{figure}

 We first consider the case with $c_{xz}=c_{yz}=0$ (we assume $x-y$ plane symmetry).
   The only non-zero $\omega_i$  is $\omega _ z = \beta_1 c_j^2 c_{xy}$.  Analysis of $c_{xy}$ equation shows a stable nontrivial fixed point for $(c_i^2)^2 >\sigma _2^2/(\beta_1\beta_2)$. Because only $\omega_z\ne 0$ the spot rotates along the equator (and so does vector $c_i$), and the particle follows a circular path.  We subsequently analyze linear stability of this solution (see \cite{SI}), by allowing modes breaking $x-y$ mirror symmetry (meaning 
 $c_{xz}$ and $c_{yz}$ non zero). A straightforward eigenvalue problem shows that  $x-y$ mirror symmetry is lost for 
 $(c_i^2)^2>3 \sigma_2^2(\beta_1\beta_2)$, and we find (besides $\omega_z$) $\omega_x= \beta_1 c_i^2 c_{yz}$ (an appropriate choice of $z$-axis allows to set $c_{xz}=0$), meaning that as soon as 
 $ c_{yz}\ne 0$, the spot moves along a circle different from equator and the particle follows a helical path (Fig. \ref{scheme3D}).   
 
 To highlight the genericity of the presented results, we write below  the evolution equations for the first and second harmonics based on symmetry only (invariance under 3D rotations), without adopting the special form (\ref{dynamicsb}) expressed by the $\beta_1$ term. To the leading order we have
\begin{subequations}
\begin{align}
\label{numdynamicsa}
	\dot c_i&=\sigma_1 c_i+\alpha_1 c_j^2c_i+\beta_1c_{ij}c_j\\
\label{numdynamicsb}
	\dot c_{ij}&=\sigma_2 c_{ij}+\beta_2(c_ic_j-\delta_{ij}c_k^2/3).
\end{align}
\end{subequations} 
Figure \ref{scheme3D} shows the full numerical result of this model which captures the analytical results.

%

{\color{black} Before concluding this section, some remarks are in order. We have written here the two harmonic equations based on symmetries. We have shown  how to derive in 2D  the two harmonic equations from an explicit phoretic model (more details can be found in \cite{FarutinC1C2}, exhibiting both qualitative and quantitative agreements with the full model). The same strategy could be adopted in 3D without additional conceptual complications.  Our goal was to highlight that two harmonics are sufficient to capture the essential features, and that symmetries can dictate the general form of the equations. The structure of the explicit phoretic model adopted here calls for an important remark, however. For the 3D version of the phoretic model adopted here, it has been shown by Rednikov et al.\cite{rednikov1994drop}  and by Morozov and Michelin \cite{Morozov2019}
that the swimming speed $V_0$ does not behave as the square root with distance from threshold (as follows from our study), but has a linear behavior. More precisely, for $Pe<Pe_1$, $V_0=0$ and $\lvert V_0\rvert \sim Pe-Pe_1$ for $Pe>Pe_1$. We note in passing that Morozov and Michelin \cite{Morozov2019} called this bifurcation {\it trancritical}, but in fact this is still  a pitchfork bifurcation, albeit non classical, since the solution $V_0=0$ becomes unstable for $Pe>Pe_1$ in favor of two symmetric solutions, $V_0\sim \pm (Pe-Pe_1)$. 
We should refer to this bifurcation as a singular pitchfork bifurcation; 
{it is definitely not a transcritical bifurcation  \cite{Morozov2019} which requires that  a fixed point branch (here the non motile state) exchanges its stability with the other fixed point branch (motile state) at their crossing junction}.  We have considered recently a simplified version of the phoretic model and found an exact analytical solution \cite{Farutin_singular}  which confirms the singular nature, in  that $\lvert V_0\rvert \sim Pe-Pe_1$ for $Pe>Pe_1$. We have shown that this singular behavior occurs only for an infinite system size (be it in 2D or 3D), whereas for a finite size (but arbitrary large) the bifurcation is a classical  pitchfork bifurcation, in that $V_0\sim ({Pe-Pe_1})^{1/2}$. This implies that our spirit of regular expansion in power series of harmonic amplitudes ($C_1$ and $C_2$) is legitimate for a finite size. In addition, we have shown that the singular behavior for infinite size is only present in the particular phoretic model presented here and considered by  Rednikov et al.\cite{rednikov1994drop}  and  Morozov and Michelin \cite{Morozov2019}. Indeed, if a slightly different version of the model is adopted \cite{Farutin_singular}, in which it is supposed that the emitted solute is, besides advection and diffusion, consumed at a certain frequency (giving rise, for example, to some product, not necessarily for interest), then the singular nature of the bifurcation for infinite systems is suppressed (even for an infinitesimal consumption rate);  the bifurcation becomes of classical pitchfork bifurcation. Nevertheless, the works of Rednikov et al.\cite{rednikov1994drop}  and  Morozov and Michelin \cite{Morozov2019} have a merit  of pointing out   a non trivial singular nature of bifurcation, rarely encountered in classical nonequlibrium systems undergoing bifurcations (such as B\'enard and Marangoni convection, Turing systems, crystal growth... which have been a focus of nonlinear community for decades). The singular nature of the bifurcation means that the radius of convergence of expansions in powers of amplitudes of harmonics goes to zero at the bifurcation point. This raises an important question of how to properly cope  a priori, for a given nonlinear model,  with the existence of   singular bifurcations in a proper manner. We have provided very recently a framework along this line \cite{Farutin_singular}.

 }

{\it Conclusion.}
The model has identified the fact that locomotory complexity  leading to diverse trajectories can be captured on the basis of symmetries and nonlinear interactions, lending evidence to its universality.
The model can be adopted for any motion fueled by a chemical field, a prominent and vast field of research  is mammalian cell motility, which is known to be  dictated by myosin and actin kinetics. 
The model can be effective not only for spherically shaped motile entities, but also 
for any shape  as long as the shape of the cell can be reconstructed from the chemical field. The model can also find application in embryonic development.
 Underlying the multicellular choreography is the actomyosin cytoskeleton dynamics, leading to the propagation 
 of localized concentration pulses\cite{Negro2019,BLANCHARD201878} affecting  cell rearrangement.
 This study also opens up new perspectives to tackle motility from a novel angle. Indeed, it should incite analytical derivation of simple nonlinear equations (as studied here) from different explicit 
 examples of motility. This will allows linking the phenomenological coefficients used here to biophysical and chemical parameters of a motile system, 
 in order to determine the conditions of  manifestation, or the lack thereof, of complex motions in parameter space, without resorting to the computationally expensive solution of the full basic model, which involves reaction-diffusion-advection with long range hydrodynamics.

We thank CNES (Centre National d'Etudes Spatiales) (C.M. S. M. R. and A.F.) for a financial support and for having access to experimental data, and the French-German university program "Living Fluids" (grant CFDA-Q1-14) (C.M., A.F. and S. R.) for a financial support 
%
\end{document}